\documentclass[12pt,a4paper]{article}
\usepackage{amsmath}
\usepackage{amssymb}
\usepackage{amsfonts}
\def \RC{{\mathbb {R}}}
\def \NC{{\mathbb {N}}}
\def \ZC{{\mathbb {Z}}}
\def \CC{{\mathbb {C}}}

\newcommand \la {\lambda}
\newcommand \bd {\begin{displaymath}}
\newcommand \ed {\end{displaymath}}

\newcommand \mf {\mathfrak}
\newcommand \pa {\partial}
\newcommand \da {\delta}

\newcommand \car {{\cal R}}
\newcommand \calp {{\cal P}}
\newcommand \calf {{\cal F}}
\newcommand \calb {{\cal B}}
\newcommand \calo {{\cal O}}
\newcommand \rto {\rightarrow}
\newcommand \dno {\downarrow}
\newtheorem{prop}{Proposition}[section]
\newtheorem{defin}{Definition}[section]

\newtheorem{lem}{Lemma}[section]
\newtheorem{thm}{Theorem}[section]
\newtheorem{corollary}{Corollary}[section]

\newtheorem{notation}{Notation}

\title{{\bf THE GENERALISED mKdV EQUATIONS\\
 FOR LEVEL $-3$ OF $\hat {{\mf s}{\mf l}}_2$}}
\author{A.Balan\\
Ecole Polytechnique\\
CMAT UMR 7640 of CNRS\\
F-91128 PALAISEAU Cedex\\
{\it email}: balan@math.polytechnique.fr}
\begin{document}
\maketitle

\abstract{A certain generalisation of the hierarchy of mKdV equations
 (modified KdV), which forms an integrable system, is studied
 here. This generalisation is based on a Lax operator
 associated to the equations, with principal components of degrees between
 $-3$ and $0$. The results are the following ones: 1) an isomorphism
 between the space of jets of the system and a quotient of
 ${Sl}_2({\CC}((t)))$; 2) the fact that the monodromy matrixes
 of the Lax operators have, morover, Poisson brackets given by the
 trigonometric $r$-matrix; 3)  a definition of the action of screening
 operators on the densities; 4) an identification of the intersection
 of the kernel with the integrals of motion.}



\newpage
\section{Introduction}
\subsection{ Foreword}

 It is treated here of certain mathematical equations,
 from the point of vue of algebra,
 as far as the jets spaces of some differential equations
 presents an isomorphism with a quotient of proalgebraic groups.  
 So here is an algebraic treatment of analytical data.
  The equations are ones of modified Korteweg-de Vries, or mKdV,
 which are generalised for level $-3$ of $\hat {{\mf s}{\mf l}}_2$.

 The Lax operator of the usual mKdV theory has terms of principal
 degrees $-1$ and $0$ in the Kac-Moody algebra $\hat {{\mf s}{\mf l}}_2$.
 The authors of \cite{F} and of \cite{GS}
 have studied analogous equations, where the Lax operator
 is a sum of components of principal degrees between $-2n-1$ and $0$;
 in the present paper, the case of the algebra
 $\hat {{\mf s}{\mf l}}_2$, $n=1$,
 is studied. The goal of the study is the classical treatment of
 the system, in the sense of a family of Poisson commutativ integrals
 of motion.

 The four following points are treated:

 1) First it is showed (theorem \ref{th1}) an isomorphism of
 proalgebraic structures between, on the one hand, a differential ring
 given by the variables of the generalised mKdV equations and
 on the other hand, the coordinates over a double quotient.
 The work is to be bound with the results of \cite{FF2}.

 2) The authors of \cite{F} and of \cite{GS} have showed Poisson brackets
 such that the generalised mKdV equations are hamiltonian systems.
 The Poisson brackets of the monodromy matrixes are calculated.
 The result (proposition \ref{croch}) is analogous
 with \cite{FT}: the Poisson brackets are given by the trigonometric
 $r$-matrix associated with $\hat {{\mf s}{\mf l}}_2$.

 3) The screening operators $Q_0$ and $Q_1$ are defined by the Poisson
 brackets of the monodomy matrix, and they can be viewed as vector fields
 over the variety of the jets. The result of the theorem \ref{trubowitz}
 is the identification with the action of ${\mf n}_+$ over $N_+/A_+$
 following from theorem \ref{th1}. The result is analogous with \cite{E}.

 4) $Q_0$ and $Q_1$ are screening operators over the density spaces. 
 In the theorem \ref{thm542}, the intersection of the kernels
 is identified with the space of the integral of motions. It is done
 by mean of a complex formed from $Q_0$ and $Q_1$ and a resolution of
 Bernstein-Gelfand-Gelfand (BGG) type.

 The problem of the quantification of the system will be studied later.   
\subsection{The presentation of the mKdV equations}

 The generalised mKdV equations are studied here
 in a matrix form.
 
 The usual mKdV equation is the following one:

\bd
 {\pa {\bf u}}/{\pa t} - 2 {\bf u}.  {\pa {\bf u}}/{\pa x} +
  {\pa^{3} {\bf u}}/{\pa x^{3}}= 0,
\ed
 with: 
\bd
{\bf u}(x,t),
\ed

 a function with real values. It is possible to consider the equation
 in the following form:

\bd
{\cal L} = \pa_{x} + p_{-1} + {\bf u} (x,t) h_{0}, 
\ed

with the matrixes:

\bd
 p_{-1}= \left( \begin{array}{ccc}
             0 & \la \\
             1  & 0 \\
         \end{array} \right),
 \tilde p_{-1}= \left( \begin{array}{ccc}
             0 & \la \\
             - 1  & 0 \\
         \end{array} \right),
 h_{0}= \left( \begin{array}{ccc}
             1 & 0\\
             0  & -1 \\
         \end{array} \right),
\ed
taking a formal parameter $\la$, so that the equation becomes:
\bd
\frac {\pa {\cal L}}{\pa t} = [{\cal L}, A],
\ed
with $A$, the following matrix:
\bd
A= p_{-3} + {\bf u} h_{-2} + {\bf u}^{2} p_{-1} + {\bf u}_{x} \tilde p_{-1}
 + ( {\bf u}_{xx} -{\bf u}^{2}) h_{0},
\ed
\bd
p_{-3}= \left( \begin{array}{cc}
               0 & \la^2\\
               \la & 0 \\
                \end{array} \right),
h_{-2}= \left( \begin{array}{cc}
             \la & 0\\
             0  & -\la \\
         \end{array} \right).
\ed

The following points are studied.
First it is possible to construct commuting flows puting the
Lax operator $\cal L$ in a conjugated form:
\bd
{\cal L} = \pa_{x} +L  = K (\pa_{x} + p_{-1} + d_1 p_1 + d_3 p_3 +
\ldots) K^{-1},
\ed
with functions $d_{2i+1}$ and:
\bd
p_{-2k+1}= \left( \begin{array}{cc}
                    0 & \la^k\\
                    \la^{k-1} & 0 \\
                   \end{array} \right),
\ed
the flows are then:
\bd
\pa_{t_{n}}{\cal L}= [ {\cal L}, {\cal A}_{n}],
\ed
with:
\bd
{\cal A}_{n} = [ Kp_{-2n-1}K^{-1}]_{-}.
\ed
Indeed, a matrix which commutes with $\cal L$ furnishes a flow
which preserves the form of $\cal L$, considering the degrees of
the terms of the bracket. The following equation is considered:
\bd
[p_{-2n-1}, \pa_{x} + \sum_{i\ge 0} d_{2i-1} p_{2i-1}]=0,
\ed
This identity is conjugated to obtain ${\cal A}_n$. So, 
a geometrical interpretation follows by mean of an isomorphism
with a quotient $N_+/A_+$; there is then 
an identification of differential rings.

A structure of Poisson brackets
over the jets is given in a synthetic way in the form:
\bd
\{ {\bf u}(x), {\bf u}(y) \} = {\frac{1}{2}}[\pa_{x}-\pa_y] \delta_{x,y},
\ed
which gives brackets for polynomial in the $\bf u$ and derivativs,
using the commutation of the bracket with the differential and the
fact that the bracket is a derivation of the functions, when
one of the sides is fixed.

It is possible to generalise the form of the operator $\cal L$
as studied by \cite{F} and \cite{GS}, taking an operator with level $-2n-1$
of $\hat {{\mf s}{\mf l}}_2$, in the form:
\bd
{\cal L} = \pa_{x} + p_{-2n-1} +L,
\ed
$L$, a matrix which depends on the levels $0,-1,-2, ..., -2n$.

Here the case of $-2n-1=-3$ is treated:
\bd
L = H_{-2} h_{-2} + E_{-1} e_{-1} + F_{-1} f_{-1} + H_{0} h_{0}.
\ed 

\subsection{Acknowledgments}

I thank greatly B.Enriquez for his help in this work, during a stay
in the Forschung-Institut f\"ur Mathematik (FIM) of Z\"urich.

\section{The isomorphism of varieties}

\subsection{ Recalls of the affine algebras}

 Let a Cartan matrix $(a_{ij})$ be, of dimension $n$ and rank $l$,
 it is possible to associate it a Kac-Moody algebra $\mf g$, called affine
 or of infinite dimension. It is given by a matrix $(a_{ij})$ and realised
 by a triplet $({\mf h}, {\Pi} , {\Pi}^{v})$,
 with $\mf h$, a complex vector space,  $\Pi=(\alpha_{i})$, in the dual
 space ${\mf h}^*$ and $\Pi^{v}=(\alpha_{i}^{v})$ in $\mf h$,
\bd
a_{ii}=2,
a_{ij} \leq 0 (\in {\ZC}_{-}),
a_{ij}=0 \Rightarrow a_{ji}=0,
\ed
 $\Pi,\Pi^v$ are linearly independant, 
\bd
 <\alpha_{i},\alpha_{j}^{v}>= a_{ij},
 <\alpha_{i},\alpha_{j}^{v}>= a_{ij},
n-l= dim ({\mf h}) -n,
\ed
 and the algebra is constructed by the generators $e_{i}$,
 $f_{i}$, $h$, with the relations:
\bd
 [e_{i},f_{j}]= \delta_{ij} \alpha_{i}^{v},
 [h, h']=0,
 [h,e_{j}]=< \alpha_{j},h>e_{j},
\ed
\bd
 [h, f_{j}]=- < \alpha_{j},h>f_{j},
ad(e_{i})^{-a_{ij}+1}(e_{j})=0,
ad(f_{i})^{-a_{ji}+1}(f_{j})=0.
\ed
Then, there is a decomposition:
\bd
{\mf g} = {\mf n}_{-} \oplus {\mf h} \oplus {\mf n}_{+},
\ed
where $\mf n_+$ and $\mf n_-$ are the sub-algebras 
constructed by the $e_{i}$ on the one hand and the $f_{j}$, on the other hand.

\medskip
\noindent

\begin{notation} .

\bd
{\mf b}_{-} =  {\mf n}_{-} \oplus {\mf h},
\ed

and:

\bd
{\mf b}_{+}=  {\mf h} \oplus {\mf n}_{+}.
\ed

\end{notation}

$\hat {{\mf s}{\mf l}}_2$ corresponds to the following Cartan matrix:

\bd
\left( \begin{array}{cc}
             2 & -2 \\
             -2 & 2 \\
         \end{array} \right).
\ed

{\notation :
\bd
h_{-2}= \left( \begin{array}{cc}
               \lambda & 0 \\
                0 & - \lambda \\
         \end{array} \right),
e_{-1}= \left( \begin{array}{cc}
             0 & \lambda \\
                0 & 0 \\
         \end{array} \right),
f_{-1}= \left( \begin{array}{cc}
             0 & 0 \\
                1 & 0 \\
         \end{array} \right),
\ed
\bd
h_{0}= \left( \begin{array}{cc}
             1 & 0 \\
                0 & -1 \\
         \end{array} \right),
p_{-3}= \left( \begin{array}{cc}
            0 & \lambda^{2} \\
              \lambda & 0 \\
         \end{array} \right),
p_{-2k+1}= \left( \begin{array}{cc}
             0 & \lambda^{k} \\
              \lambda^{k-1} & 0 \\
         \end{array} \right),
\ed

\bd
\tilde p_{-2k+1}= \left( \begin{array}{cc}
             0 &  -\la^{k} \\
              \la^{k-1} & 0 \\
         \end{array} \right),
h_{2n}= \left( \begin{array}{cc}
               \la^{-n} & 0 \\
              0 & - \la^{-n} \\
         \end{array} \right),
\ed
\bd
i_{2n}= \left( \begin{array}{cc}
               \la^{-n} & 0 \\
                0 & \la^{-n} \\
         \end{array} \right),
\ed
\bd
e_{2n+1}= \left( \begin{array}{cc}
             0 & \la^{-n} \\
                0 & 0 \\
         \end{array} \right),
f_{2n+1}= \left( \begin{array}{cc}
             0 & 0 \\
                \la^{-n-1} & 0 \\
         \end{array} \right).
\ed}

 The set of the $2 \times 2$ matrixes with 
 coefficients in the ring $\CC[[\la^{-1}]]$, 
 with $1$ determinant and $val(m_{12}) \ge 1$ is noted $N_{+}$.

 The set of matrixes, exponential of the sub-algebra generated
 by the $p_{2k+1}$ is noted $A_+$.

\medskip
\noindent

\subsection{The isomorphism of differential rings}

\medskip
\noindent

In this section, the following theorem is treated:

\medskip
\noindent

\begin{thm} . \label{th1}
 It exits an isomorphism of differential rings
 between the one of the jets of the variables of the
 generalised mKdV equations, with the canonical differential and the
 one of the coordinates over the quotient $N_{+}/A_{+}$, with the 
 differential given by the right action of:
\bd
p_{-3} + d_{-1} p_{-1},
\ed
over $N_+/A_+$.
\end{thm}

\medskip
\noindent

 First is established an isomorphism of rings, and then are identified
 the differentials.
\subsection{Determination of a conjugation}
 In the generalised mKdV equations, the operator $\cal L$ is put
 in a conjugated form and the unicity of the conjugation is showed, \cite{DS}
 provided that a constraint is imposed over the matrix of passage.
\subsection{The Lax operator in a conjugated form}
 The Lax operator of the generalised mKdV equations is:
\begin{equation} \label{1.1}
 {\mathcal L}= \pa_{z} + p_{-3} + H_{-2} h_{-2} + E_{-1} e_{-1}
 + F_{-1} f_{-1} +H_{0}h_{0}. 
\end{equation}
 With $z \in {\RC}$
 and $E_{-1},F_{-1},H_{-2},H_{0}$, some smooth functions of $z$.

\medskip
\noindent

{\defin .
 For a  matrix of rank $n$, $A=(a_{i,j})$,
 an {\it antitrace} is defined (sum of the terms over the antidiagonal):
\bd
\tau (A)= \sum_{i=0}^{n-1} a_{n-i,i+1}.
\ed}

 A conjugated form of the operator
 $\cal L$ in the generalised mKdV equations is then presented.

\medskip
\noindent

{\prop . \label{conj}
 The Lax operator has the following conjugated form:
\bd
 {\cal L} = M ( \pa_{z} + p_{-3} + d_{-1}p_{-1} +
\ed
\begin{equation}
d_{1} p_{1} + \sum_{i\ge -1} d_{2i+1}p_{2i+1} ) M^{-1}. \label{1.2}
\end{equation}}

 The $d_{2i+1}$ are polynomials over the
 variables $E_{-1},F_{-1},H_{0},H_{-2}$ and derivatives, 
\footnote{This set is noted $\pi_0$.} and the matrix $M$ 
 belongs to $M_{2}( {\pi}_0 [[\la^{-1}]])$, the $2 \times 2$ matrixes, with 
 coefficients in the ring $\pi_0[[\la^{-1}]]$, is uniquely defined,
 with $1$ determinant, $val(m_{12}) \ge 1$,
 \footnote{This set is noted $N_{+}({\pi}_0)$.}
 and with $\tilde \tau(M)=0$.

\subsection{ Appendix}

 There is a property for the adjoint action $ad(p_{-3}))$, linear application
 of  ${\mf s}{\mf l}_2((\la))$, a decomposition in a direct sum:
\bd
{\mf s}{\mf l}_2((\la))= Ker (ad(p_{-3})) \oplus Im(ad(p_{-3})) ,
\ed
$Ker (ad(p_{-3}))$ is defined as the Lie algebra whose base is formed by
 $p_{2k+1}$, and $Im(ad(p_{-3}))$, the
 $h_{2i}$ and the $\tilde p_{2i+1}$. $ad(p_{-3})$ is an isomorphism
 over its image.

\subsection{Isomorphism of differential rings}

 An isomorphism is showed between $\pi_0^c=\pi_{0}/[d_{-1}=c]$
 and the ring of coordinates over the quotient $N_{+}/A_{+}$, 
$N_{+}$ is the group associated with positiv nilpotents
 of the algebra of infinite dimension ${\mf s}{\mf l}_{2}((\la))$,
 $A_{+}$ is the canonical commutativ sub-group.

 The differential ring $\pi_0={\CC}[Jets]$ is defined
 as the ring of polynomials given by the variables,
\bd
 X_k = \{  E_{-1;k}=E_{-1}^{(k)} , F_{-1;k}=F_{-1}^{(k)},
 H_{0;k}=H_{0}^{(k)}, H_{-2;k}=H_{-2}^{(k)}, k \geq 0 \},
\ed
 and the differential given by the derivation  $\pa$.
\bd
\pi_{0}= {\CC} [Jets]= 
\ed
\bd
{\CC} [E_{-1;k},F_{-1;k},H_{0;k},H_{-2;k}, k
\in \NC ].
\ed

 For $c$ a complex number, the differential ring $\pi_0^c$ is defined as:  
\bd
\pi_0^c= {\CC}[Jets_c]=
\ed
\bd
{\CC}[E_{-1;k},F_{-1;k},H_{0;k},H_{-2;k} ; k
\in \NC ] / (d_{-1}-c, d_{-1}^{(k)}), 
\ed
it is clear that $\pa$ induces a derivation over ${\CC}[Jets_c]$. 

The elements of the rings ${\CC}[Jets]$ and ${\CC}[Jets_c]$ are 
respectivly  functions over the spaces: 
\bd
Jets = \{ (E_{-1;k}, F_{-1;k},H_{0,k},H_{-2;k}) \in ({\CC})^4, k \in \NC \},
\ed
and:
\bd
Jets_c = \{(E_{-1;k},F_{-1;k},H_{0,k},H_{-2;k}) \in ({\CC})^4, k \in \NC  | 
(d_{-1} - c)^{(k)} = 0 \}. 
\ed

{\defin  .
\bd
J= \{ M \in M_{2}({\CC}(({\la}^{-1})))/ det (M) = 1, \tilde \tau (M)=0 \},
\ed
  the $2 \times 2$ matrixes, with $1$ determinant and
$\tilde \tau=0$; these conditions being algebraic,
 an algebraic sub-variety is obtained.}

{\lem  . 
A natural application from $J$ to $N_+/A_+$ is an isomorphism of
 proalgebraic varieties.}

{\corollary .
 So:
\bd
N_{+}=J.A_{+}.
\ed}

\medskip
\noindent

{\thm  .
 For all complex number $c$, the decomposition which associates
 to all jet of functions $E_{-1},F_{-1},H_{-2},H_{0}$,
 an element $M$ of $N_+$, induces an isomorphism of 
 proalgebraic varieties between  $Jets_c$ and
  $N_+/A_+$:
\bd
 {\CC}[Jets_c] \cong {\CC}[N_+/A_+]. \label{iso}
\ed}

{\corollary .  \label{iso:total}
The application of $Jets$ toward: 
\bd
N_+/A_+ \times {\CC}^{\infty},
\ed 
associating to a jet the couple formed by the class of $n_+$ and 
the family $(d_{-1}^{(k)})_{k\ge 0}$, defines an isomorphism
 of differential rings between $({\CC}[Jets] , \pa)$ and: 
\bd
{\CC} [N_{+}/A_{+}] \otimes {\CC} [ x_0,x_1,... ],
\ed
with the differential: 
\bd
r(p_{-3} + x_{0}p_{-1}) \otimes 1 + 1 \otimes (\sum_{i\ge 0} x_{i+1} 
{\pa / {\pa x_i}}),
\ed
$r$ is the right regular action.}

 An isomorphism between the rings of infinite dimension, the jets of
 variables in the generalised mKdV equations and the coordinates
 of a quotient has been showed.

\subsection{The differential for the quotient}

 In the generalised mKdV equations, an identification
 of the differential for the quotient is here showed. 
 The differential over the jets is the derivation of the variables.

\medskip
\noindent

{\defin .
  Let $\calo$ be, the matrixes of zero trace, of $-\la$ determinant,
 and with positiv terms in $\la^{-1}$ and with $\la$ too:
\bd 
\left( \begin{array}{cc}
               a & b\\
               c & -a \\            
                 \end{array} \right),
\ed 
with the determinant of the matrix $-\lambda$, which gives:
\bd
a^{2} + bc = \lambda.
\ed}

\medskip
\noindent

{\lem .
 An identification between $N_{+}/A_{+}$ and ${\calo}$ is given by the arrow:
\bd
N_{+}/A_{+} \to {\calo},
\ed
\bd
 M \mapsto {\cal V}=Mp_{-1}M^{-1}.
\ed}

\medskip
\noindent
The right action of the  $p_{2i+1}$ over ${\calo}$ is:

\bd
[(\la^{-i-1} {\cal V})_{-},{\cal V}].
\ed

{\prop .
 The differential of the ring of jets of the variables is given
 by the right action over the quotient  $N_{+}/A_{+}$, of: 
\bd
p_{-3}+ d_{-1}p_{-1}.
\ed}

\medskip
\noindent

{\bf Formula}:

\medskip
\noindent
\bd
(\la {\cal V})_{-} + c ({\cal V})_{-} = p_{-3}+L,
\ed
avec:
\bd
c= ( E_{-1}+F_{-1}+H_{-2}^{2})/2.
\ed

 Doing the study of the generalised mKdV equations,
 an isomorphism of proalgebraic structures has been showed
 and, too, an identification of two differential rings, on the one hand,
 the ring of the jets and derivatives and in the other hand, the ring
 of the coordinates over a quotient with differential given by a right action.
 The generalised mKdV equations have so been studied in link with the affine 
 algebra
 ${\mf s}l_{2}((\la))$. So, an isomorphism of differential rings has been
 established. 

\section{The structures of Poisson brackets}

\subsection{The Poisson brackets over the variables}
 The Poisson brackets for the generalised mKdV equations 
 have been defined by the authors of \cite{F} in the following way:
\bd
{\cal R} = {1 / 2}[\calp_{\alpha} - \calp_{\beta}],
\ed
the half difference of the projection, with $\alpha$ and $\beta$,
 the spaces of the affine algebra $\hat {{\mf s}{\mf l}}_2$ corresponding with
 ${\mf b}_-$ and ${\mf n}_+$. The Poisson brackets are:
\bd
\{ f, g \}_{\car} (L) =  \int_{- \infty}^{+ \infty} dx \langle L ,
 [ ({{\delta f} / {\delta L}}) ,
 ({{\delta g} / {\delta L}})]_{\car} \rangle
 -
\ed
\bd
 \langle \car  ({ {\delta f} / {\delta L}}) , \pa_x 
 ({{\delta g} / {\delta L}}) \rangle - 
 \langle \car  ({{\delta f} / {\delta L}}) , \pa_x \car 
 ({{\delta g} / {\delta L}}) \rangle.
\ed
\bd
\langle u, v \rangle,
\ed
 the bilinear form in the affine algebra for two elements
 $u$ and $v$ of $\hat {\mf sl}_2$,
\bd
{\cal L} = \pa_z + L,
\ed
the Lax operateur for the generalised mKdV equations,
\bd
f,g,
\ed
 two functionals over the space of jets of the variables
 in the generalised mKdV equations, (as for example
 $\int_{-\infty}^{+\infty} E_{-1}(x) dx$
 \footnote{Take variables with compact support.}, or more generaly,
 the integral of a  polynomial in the jets of the variables).
 It is a way to give Poisson brackets over the jets
 of the variables of the generalised mKdV equations.
 So the brackets have been obtained,
 it should be too possible to give them in an equivalent way
 over the jets of variables, or over the monodromy matrix.

\subsection{The structures of Poisson brackets over the jets}

 The brackets of the variables in the generalised mKdV equations
 are given in the following way:

\medskip
\noindent
{\prop . 
 Over the jets of the variables, the Poisson brackets are:}
\bd
\{ H_0 (x), H_0 (y) \} = \frac{1}{2} \pa_{x}\da_{x,y},
\ed
\bd
\{ H_0 (x), E_{-1} (y) \} = 0, 
\{ H_0 (x), F_{-1} (y) \} = 0,
\{ H_0 (x), H_{-2} (y) \} = 0, 
\ed
\bd
\{ E_{-1} (x) , E_{-1} (y) \} =0,
\{ F_{-1}(x) , F_{-1}(y) \} =0,
\ed
\bd
\{ E_{-1}(x), F_{-1}(y) \} = -4 H_{-2}(x) \da_{x,y},
\ed
\bd 
\{ E_{-1}(x), H_{-2}(y) \} = 2 \da_{x,y},
\{ F_{-1}(x), H_{-2}(y) \} = - 2 \da_{x,y},
\ed
\bd
\{ H_{-2}(x), H_{-2}(y) \} = 0.
\ed
 There is moreover associativity of the Poisson brackets in the variables.

\subsection{$d_{-1}$ central element}

\medskip
\noindent
{\prop .
\bd
 d_{-1}(x) = {1 / 2}[E_{-1}(x) + F_{-1}(x) + H_{-2}^2(x)],
\ed
 is a central element for the Poisson bracket.}

\section{The Poisson brackets of the monodromy matrix}

 The matrix for the monodromy is
 $Pexp \int_a^b L(x) dx$, with $L=p_{-3} + H_0 h_0
+ H_{-2} h_{-2} + E_{-1} e_{-1} + F_{-1} f_{-1}$.
There is so a way to found Poisson brackets of two functions,
 $f,g$, at the level of the monodromy matrix from the synthetic formula.
 The following bracket has to be computed:
\bd
\{ Pexp \int_a^b L(x) dx \stackrel {\otimes} {,} Pexp \int_a^b L(y) dy \}.
\ed
Let $r_{trigo}$ be: 
\bd
r_{trigo} = [\la + \mu]/[\la-\mu] t + e \otimes f -f \otimes e ,
\ed
\bd
t= 1/2 h \otimes h + e \otimes f + f \otimes e.
\ed
\medskip
\noindent
{\prop . \label{croch}
 At the level of the monodromy matrixes, the Poisson brackets
 are given by the following formula:
\bd
\{ Pexp \int_a^b L(x) dx \stackrel {\otimes} {,} Pexp \int_a^b L(y) dy \} =
\ed
\bd
[r_{trigo}, [Pexp \int_a^b L(t) dt] \otimes  [Pexp \int_a^b L(t) dt]].
\ed} 

\section{ The action of $Q_0$ and of $Q_1$ over the jets of the variables
 generalised mKdV}

 The screening operators are the following integrals:

\bd
Q_0  = \int_{\RC}  E_{-1}(u) e^{2 \int_{- \infty}^u H_0 (v) dv} du,
\ed
\bd
Q_1  = \int_{\RC} F_{-1}(u) e^{- 2 \int_{- \infty}^u H_0 (v) dv} du.
\ed

\subsection{The action of $\bar Q_0$ and $\bar Q_1$ over the variables}

 In a heuristic way, the action of  $\bar Q_0$ and $\bar Q_1$
 over the polynomials  in the jets is the following:
\bd
\bar Q_0 (P)(u) = e^{2 (\int_{-\infty}^u H_0 (v) dv)} \{ Q_0 , P \}(u),
\ed
with $P$, a polynomial in the jets of the variables of the
 generalised mKdV equations.
 The actions of $\bar Q_0$ and $\bar Q_1$ can be written down
 as vector fields:
\bd
\bar Q_0 = \sum_{n \in \NC} a_n {\pa  / {\pa H_0^{(n)}}}+
 b_n  {\pa  / {\pa E_{-1}^{(n)}}}+
 c_n  {\pa  / {\pa F_{-1}^{(n)}}}+
 d_n  {\pa  / {\pa H_{-2}^{(n)}}}.
\ed

{\notation .
\bd
h_0^{[n]} = {\frac { H_0^{(n)}}{n!}},
e_{-1}^{[n]} = {\frac {  E_{-1}^{(n)}} {n!}},
f_{-1}^{[n]} = {\frac {  F_{-1}^{(n)}}{n!}},
h_{-2}^{[n]} = {\frac {  H_{-2}^{(n)}} { n!}}.
\ed}

\medskip
\noindent
{\prop .
 The action of $\bar Q_0$ is given by the following series:
\bd
\sum_{n \in \NC} a_n t^n= -
(\sum_{n \in \NC} e_{-1}^{[n]} t^n)( e^{-2 \sum_{n \in \NC} h_0^{[n]}
{ t^{n+1} / (n+1)}} ),
\ed
\bd
\sum_{n \in \NC} b_n t^n =0,
\ed
\bd
\sum_{n \in \NC} c_n t^n = 4
(\sum_{n \in \NC} h_{-2}^{[n]} t^n)( e^{- 2 \sum_{n \in \NC} h_0^{[n]}
{ t^{n+1} / (n+1)}} ),
\ed
\bd
\sum_{n \in \NC} d_n t^n = -  2
e^{2 \sum_{n \in \NC} h_0^{[n]}
{ t^{n+1} / (n+1)}} .
\ed}

\medskip
\noindent
{\thm .
 The series for $\bar Q_1$ are:
\bd
\sum_{n \in \NC}   a'_n =
 (\sum_{n \in \NC} e_{-1}^{[n]} t^n)( e^{ 2 \sum_{n \in \NC} h_0^{[n]}
{ t^{n+1} / (n+1)}} ),
\ed
\bd
\sum_{n \in \NC}   b'_n = - 4
(\sum_{n \in \NC} h_{-2}^{[n]} t^n)( e^{ 2 \sum_{n \in \NC} h_0^{[n]}
{ t^{n+1} / (n+1)}} ),
\ed
\bd
\sum_{n \in \NC}   c'_n = 0,
\ed
\bd
\sum_{n \in \NC}   d'_n =  2
e^{2 \sum_{n \in \NC} h_0^{[n]}
{ t^{n+1} / (n+1)}}. 
\ed}

\subsection{The action of the screening operateurs}

 Here is a result over the action of the $\bar Q_i$,
 which is identified with the one of
 $\mf n_+$ over the quotient, according to the following theorem:

\medskip
\noindent
{\thm . \label{trubowitz}
 The action of $\bar Q_0$ and $\bar Q_1$ is defined by the Poisson brackets.
 With the isomorphism, the action is identified as being the one
 of $\mf n_+$ over the coordinates of the quotient
 ${\CC} [ N_+/A_+]$.}

 First the action of the $Q_i$ is identified over the space  $\cal O$
 (matrixes of $-\la$ determinant and zero trace). Then, it is showed
 a lemma concerning the inclusion of the ring $R_i$ 
 in the kernel of the operator $\bar Q_i$, next an other lemma over the
 coefficients of ${\cal V}_i$,
 and last the theorem follows.
\section{Identification of the action of $\bar Q_i$ over the ${\cal V}_i$}
 In the generalised mKdV equations, the action of the
$\bar Q_0$ and $\bar Q_1$ is identified with help of a conjugation
 by the operators:
\bd
{\cal L}^{(0)} = \pa_x + L^{(0)},
{\cal L}^{(1)} = \pa_x + L^{(1)},
\ed
two operators defined as being:
\bd
L^{(0)}= p_{-3} +
E_{-1}^{(0)} e_{-1} + F_{-1}^{(0)}f_{-1} + H_{0}^{(0)} h_{0} +
F_{1}^{(0)} f_1,
\ed
\bd
L^{(1)}= p_{-3} + 
E_{-1}^{(1)} e_{-1} + F_{-1}^{(1)}f_{-1} + H_{0}^{(1)} h_{0} +
E_{1}^{(1)} e_1,
\ed
these operators are conjugable with ${\cal L}$, the Lax operator
 of the generalised mKdV equations.
\bd
{\cal L}^{(0)} = n_0 ( \pa_x + L) n_0^{-1},
{\cal L}^{(1)} = n_1 ( \pa_x + L) n_1^{-1},
\ed
with:
\bd
L = p_{-3} + H_{-2} h_{-2} + E_{-1} e_{-1} + F_{-1} f_{-1} + H_0 h_0.
\ed
The matrixes of conjugation are the following ones:
\bd
n_1 = exp ( - H_{-2} e_1),
n_0 = exp ( H_{-2} f_1).
\ed

There is then an identification of the action over $\cal O$;
 the action of $\bar Q_0$ and $\bar Q_1$ is identified over the matrixes
  $\cal V$ in $\cal O$ (the matrixes of $1$ determinant and zero trace).
\bd
{\cal V}^{(0)} = (n_0 n_+ ) p_{-1} (n_0 n_+)^{-1},
{\cal V}^{(1)} = (n_1 n_+) p_{-1} (n_1 n_+)^{-1}. 
\ed

\subsection{Inclusion of $R_i$ in $Ker \bar Q_i$}
$R_0$ et $R_1$ are defined as being the differential rings generated
 by the coefficients of 
 ${\cal L}^{(0)}$ and of ${\cal L}^{(1)}$.

\medskip
\noindent
{\lem .
 $R_i$ is contained in $Ker \bar Q_i$.}

\subsection{Lemma for the coefficients of ${\cal V}_i$}

 Then, for $i = 0,1$, the coefficients of ${\cal V}_i$ are in $R_i$.
 To see it, the following result must be showed:

\medskip
\noindent
{ \lem .
 The following conjugations hold:\label{trubo}
 
\begin{equation} \label{trub}
{\cal L}_i = K_i( \pa_x + p_{-3} + \sum_{j \ge 0} d_{2j-1}^{(i)}
 p_{2j-1}){K_i}^{-1}, 
\end{equation}
where the $K_i$ are matrixes of $Sl_2(R_i((\la)))$ and the 
$d_{2j-1}^{(i)}$ belong to $R_i$. }

The fact that the coefficients of ${\cal V}_i$ are in $R_i$ can be showed
 then in the following way. By definition of the ${\cal L}_i$:
\bd
{\cal L}_i = (n_i n_+) (\pa + \sum_{j\ge 0} d_{2j-1}^{(i)}
 p_{2j-1}) (n_i n_+)^{-1}. 
\ed
Comparing the formula with (\ref{trub}), it is possible to deduce
 that the classes of $K_i$ and of $n_i n_+$ in $N_+/A_+$ are the same. 
It follows the equality of
 $K_i p_{-1} K_i^{-1}$ and of  
${\cal V}^{(i)}$.
 From the lemma \ref{trubo}, the coefficients of $K_i$ are in $R_i$,
 which implies that the coefficients of 
${\cal V}^{(i)}$ are in $R_i$ too. 

\subsection{The action of $\bar Q_0$ and $\bar Q_1$}

{\thm .
$\bar Q_0$ and $\bar Q_1$ define vector fiels over the space
 $Jets_c$, which, by the isomorphism between the jets and the
 coordinates over $N_+/A_+$, are
 identifiable with the regular action
 of the generators $2f_{-1}$ and $2e_{-1}$ of $\mf n_+$ over $N_+/A_+$.}

\section{The intersection of the kernels of $\tilde Q_i$}

 The intersections of the kernels of the operators
 $\tilde Q_i$, defined just above, are here identified with
 the integrals of motion given by the $d_{2i+1}$.
 To obtain the result, a resolution is used, of Bernstein-Gelfand-Gelfand
 (BGG) type of the modulus $\pi_0$ \cite{FF} in the jets
 over which $\mf n_+$ acts.

\subsection{Recalls of  cohomology}
The cohomology for the algebra  ${\mf n_+}$ with coefficients in the modulus
 $M$ is defined by:
\bd
H^*({\mf n}_+, M) = Hom (R^*, M),
\ed
with $R^*$, a resolution of ${\mf n}_+$. It is possible to find one
 by mean of the following complex:
\bd
C^*(M)= \{ \Psi : \Lambda^* {\mf n}_+ \rto M \},
\ed
which has the differential:
\bd
\pa \Psi (x_1, x_2, \ldots , x_{n+1})
 =
\ed
\bd
 \sum_{ i < j }
 (-1)^{i+j+1} \Psi ([x_i,x_j], x_1  ,  x_2, \ldots  , \check x_i, 
\ldots , \check x_j, \ldots, x_{n+1}) +
\ed
\bd
+ \sum_{i=1}^{n+1}  x_i. \Psi (x_1, x_2, \ldots , \check x_i, \ldots , 
x_{n+1}) .
\ed 
It is possible to verify:
\bd
\pa_{n,n+1} \circ \pa_{n-1,n} =0,
\ed
\bd
H^0({\mf n_+}, M) = \{ m, n.m =m, \forall n \in {\mf n}_+ \},
\ed  
\bd
H^1({\mf n_+}, M) = Z^1({\mf n_+}, M)/B^1({\mf n_+}, M),
\ed
\bd
Z^1({\mf n_+}, M) = \{ \Psi: {\mf n_+} \rto M, \Psi([x,y])
 =  x \Psi(y) - y \Psi(x) \}, 
\ed
\bd
B^1({\mf n_+}, M) = \{ \Psi:  {\mf n_+} \rto M, \exists m \in M,
 \Psi(x) = x.m \}. 
\ed
\subsection{The BGG resolution}

The general BGG resolution, for a modulus $M$ over the algebra 
 ${\mf n}_+$, is the following complex:
\bd
\calf^* (M), \calf^0 = M, \calf^1 = \calf^2 = \ldots = M^{\oplus 2},
\ed
with the differentials $d_{01}$, $d_{12}$, $d_{i,i+1}$,
\bd
d_{01}= e_0 \oplus e_1,
\ed
\bd
 d_{12}|_{M \times 0} = ( e_0^3, -3e_0^2 e_1 + 3e_0 e_1 e_0+
e_1 e_0^2),
\ed
\bd
 d_{12}|_{0 \times M}= ( e_0e_1^2-3e_1e_0e_1+3e_1^2e_0,e_1^3),
\ed
\bd
d_{i, i+1}|_{M \times 0}= ( e_0^{2i+1}, p_i  ),
 d_{i, i+1}|_{0 \times M} = ( q_i  , e_1^{2i+1}),
\ed
with $p_i, q_i \in U {\mf n}_+$, 
so that the applications form a complex. 

\medskip
\noindent
{\thm \cite{BGG} .
The cohomology of the complex is: 
\bd
H^*( {\mf n}_+, M).
\ed}

\subsection{Lemma over the kernel of the operators $\tilde Q_i$}

In the case of the modulus $\pi_0= \CC [Jets]$, and
 $\pi_0^c= \CC [Jets_c]$, with action of ${\mf n}_+$, 
 the $\pi_n$ are defined as being the modulus $\pi_0$,
 with a differential $\pa + 2n H_0$, and the
 $\tilde \pi_n = \pi_n / Im (\pa_n)$.
The operators $\bar Q_i$ and $\tilde Q_i$, act over them, by mean of the
 commutation lemma:
\bd
\bar Q_0 \circ \pa_n = \pa_{n-1} \circ \bar Q_0,
\ed
\bd
\bar Q_1 \circ \pa_n = \pa_{n+1} \circ \bar Q_1.
\ed
 Then, the bicomplex $\calb^*(\tilde \pi_0, \da)$, ($\da$, the differential)
 is considered:
\bd
\calf^* \stackrel {\pa_*} {\rto} \calf^*,
\ed
$$
\begin{array}{cccccc}
 \calf^0  & \stackrel {\pa_0} {\rto} &  \calf^0 & \\\
\dno & & \dno &  \\
 \calf^1 & \stackrel {\pa_1} {\rto} & \calf^{1} & \\
\dno & & \dno &  \\
  . & . & . \\
  . & . & . \\
  . & . & .\\
\dno & & \dno &  \\
 \calf^n & \stackrel {\pa_n} {\rto} & \calf^{n} & \\
\dno & & \dno &  \\
 .       & .                               & .         & \\
 . & . & . &\\
 . & . & . &\\
\end{array}
$$

\medskip
\noindent
{\lem .
\bd
 \cap_{i= 0,1} Ker (\tilde Q_i) = H^1 (\calb^*(\tilde \pi_0, \da)).
\ed}

\subsection{The kernels of $\tilde Q_i$, integrals of motion}

 A definition of a graduation over the space of the jets
 and the coordinates of the algebraic quotient is given.
 Over $\CC [N_+/A_+]$, the following graduation is defined: for
$V = \oplus_i V_i$ a ${\mf n}_+$-modulus of graduation compatible
 with the principal one of ${\mf n}_+$, and $V^*$, its dual;
 it is graduated by  $(V^*)_i = (V_{-i})^*$. For $v$ in $V$, annulated by 
${\mf a}_+$, $\Psi$ in $V^*$ and  $n$ in ${\mf n}_+$, it is defined:

\bd
{n} \mapsto \langle \Psi, {n} v \rangle \in \CC [N_+/A_+],
\ed
 the degree of the function defined equal to
$i-j$. It defines a structure of graduated ring over $\CC[N_+/A_+]$. 
 A natural graduation over the cohomology $H^i( {\mf n}_+, \CC[N_+/A_+])$ 
 can be deduced then. The isomorphism of the Shapiro lemma is compatible
 with the graduation. By the isomorphism with
$\cal O$, the coefficients of the matrix ${\cal V}_i$ are of degree $-i-1$;
 it can be showed by application of the above definition to the adjoint
 representation of ${\mf n}_+$. On the other hand, it is demonstrated
 by reccurence that the
 $d_{2i+1}$ are of degree $2i+1$.
Let the degree over $\CC[Jets]$ be: 
\bd
deg(H_{-2})= -1,
deg(E_{-1})= deg(F_{-1})=-2, \label{deg:1}
\ed
\bd
deg(H_0)=-3,
deg(\pa) = -3.
\ed

The differential ring  $(\CC[d_{-1}^{(i)}] , \pa_0)$ has
the degree defined by:
\bd
deg(d_{-1}) = -2, deg(\pa_0) = -3,
\ed
and the ring 
$\CC [N_+/A_+] \otimes \CC [ d_{-1}^{(i)}]$, the degree defined 
 as the product degree with the one of \ref{deg:1}. 

\medskip
\noindent
{\lem .
 The isomorphism:
\bd
 \CC [Jets] \cong \CC [N_+/A_+] \otimes \CC [ d_{-1}^{(i)}],  
\ed
established in corollary \ref{iso:total}, preserves the degrees.} 

 The following theorem concerns then the case $\pi_0$ and $\pi_0^c$ too:

\medskip
\noindent
{\thm . \label{thm542}
 The intersection of the kernels of the operators
 $\tilde Q_i$ is the vector 
 space generated by the integrals of the $d_{2i+1}$
 \footnote{The $d_{2i+1}$ have been defined in the proposition \ref{conj}
 for the dressing of Drinfeld-Sokolov \cite{DS}.}:
\bd
\oint d_{2i+1}(x) dx,
\ed
 in the case of $\pi_0^c$, and in the case of $\pi_0$, too by the
 functionals:
\bd
\oint P(d_{-1}, d_{-1}', ..., d_{-1}^{(n)}, ...)(x) dx.
\ed}

\medskip
\noindent
 
 The $d_{2i+1}$ are morover elements which are in involution in the algebra
 of the jets fited with the Poisson brackets.

\medskip
\noindent

{\prop . The $\oint d_{2i+1}$, which are in $\CC [ \pi_0^0]$, are in involution
 for the Poisson brackets.}

\end{document}